\documentclass[%
 reprint,
superscriptaddress,
 amsmath,amssymb,
 aps]{revtex4-2}

\usepackage{graphicx}% Include figure files
\usepackage{dcolumn}% Align table columns on decimal point
\usepackage{bm}% bold math
\usepackage[version=3]{mhchem} % chemical formula
\usepackage{xcolor} % text coloring
 
 \usepackage{ulem}
 \usepackage{balance}

\begin{document}

\preprint{APS/123-QED}

\title{High-harmonic generation in liquids with few-cycle pulses: effect of laser-pulse duration on the cut-off energy}%Force line breaks with \\

\author{Angana Mondal}
\affiliation{Laboratorium für Physikalische Chemie, ETH Zürich, Zurich, Switzerland}  

\author{Benedikt Waser}
\affiliation{Laboratorium für Physikalische Chemie, ETH Zürich, Zurich, Switzerland}  

\author{Tadas Balciunas}

\affiliation{Laboratorium für Physikalische Chemie, ETH Zürich, Zurich, Switzerland}

\author{Ofer Neufeld}
\affiliation{Max Planck Institute for the Structure and Dynamics of Matter, Luruper Chaussee 149, 22761 Hamburg, Germany}
\affiliation{Center for Free-Electron Laser Science CFEL, Deutsches Elektronen-Synchrotron DESY, Notkestra\ss e 85, 22607 Hamburg, Germany}

\author{Zhong Yin}
\affiliation{Laboratorium für Physikalische Chemie, ETH Zürich, Zurich, Switzerland}

\author{Nicolas Tancogne-Dejean}
\affiliation{Max Planck Institute for the Structure and Dynamics of Matter, Luruper Chaussee 149, 22761 Hamburg, Germany}
\affiliation{Center for Free-Electron Laser Science CFEL, Deutsches Elektronen-Synchrotron DESY, Notkestra\ss e 85, 22607 Hamburg, Germany}

\author{Angel Rubio}
\affiliation{Center for Free-Electron Laser Science CFEL, Deutsches Elektronen-Synchrotron DESY, Notkestra\ss e 85, 22607 Hamburg, Germany}
\affiliation{Max Planck Institute for the Structure and Dynamics of Matter, Luruper Chaussee 149, 22761 Hamburg, Germany}
\affiliation{Center for Computational Quantum Physics (CCQ), The Flatiron Institute, 162 Fifth Avenue, New York NY 10010, USA}

\author{Hans Jakob W\"orner}
\email{hwoerner@ethz.ch}
\affiliation{Laboratorium für Physikalische Chemie, ETH Zürich, Zurich, Switzerland}

\begin{abstract}
High-harmonic generation (HHG) in liquids is opening new opportunities for attosecond light sources and attosecond time-resolved studies of dynamics in the liquid phase. In gas-phase HHG, few-cycle pulses are routinely used to create isolated attosecond pulses and to extend the cut-off energy. Here, we study the properties of HHG in liquids, including water and several alcohols, by continuously tuning the pulse duration of a mid-infrared driver from the multi- to the sub-two-cycle regime. Similar to the gas phase, we observe the transition from discrete odd-order harmonics to continuous extreme-ultraviolet emission. However, the cut-off energy is shown to be entirely independent of the pulse duration. This observation is confirmed by {\it ab-initio} simulations of HHG in large clusters. Our results support the notion that the cut-off energy is a fundamental property of the liquid, independent of the driving-pulse properties. Combined with the recently reported wavelength-independence of the cutoff, these results confirm the direct sensitivity of HHG to the mean-free paths of slow electrons in liquids. Our results additionally imply that few-cycle mid-infrared laser pulses are suitable drivers for generating isolated attosecond pulses from liquids.
\end{abstract}

\maketitle

%\tableofcontents

%%%%%%%%%%%%%%%%%%%%%%%%%%  body  %%%%%%%%%%%%%%%%%%%%%%%%%%
\section{Introduction}
The process of high-harmonic generation (HHG) has been successfully used as a probe method for understanding strong-field dynamics in gases and solids with inherent attosecond time resolution. This has led to the possibility of imaging and reconstruction of molecular orbitals \cite{itatani04a,haessler10a,peng2019}, time-dependent chirality \cite{baykusheva18a,baykusheva2019}, inter-band and intra-band electron dynamics \cite{Ghimire2011,vampa2015,luu2015,luu18b,ghimire2019} and probing of charge migration \cite{Kraus2015,woerner17a,kraus18a}, to name a few. The basis of such applications of HHG to high-harmonic spectroscopy (HHS) is a clear understanding of the HHG mechanism. This has been achieved in gases through systematic studies of fundamental observables like the cut-off energy as a function of laser parameters, such as intensity, wavelength, pulse duration, etc. \cite{Huillier1993, gordon2005,pulsewidth1,Yun_2010, Schmidt_2012}. However, most (bio)chemically-relevant reactions occur in the liquid phase. As a result, to apply HHS to liquids, it is not only important to understand the HHG mechanism in liquids but also to systematically investigate the harmonic spectrum dependence on various experimental parameters for identifying features that are laser driven and those that originate from structural properties of the liquid environment.

Previous works on HHS in bulk liquids have been demonstrated to be a bright source of extreme-ultraviolet radiation \cite{Luu2018, Yin_2020, svoboda2021polarization}. Further, in the multi-cycle regime, the cut-off energy $E_{\rm c}$ of the liquid harmonic spectra as defined by Lewenstein et al. \cite{Lewenstein1994}, i.e. the end of the plateau region, has been shown to be independent of wavelength and the laser intensity \cite{Wavelength2022}. These results have been successfully explained within a scattering-limited trajectory model that shows the harmonic cut-off energy to be limited by a sample-characteristic mean-free path \cite{Wavelength2022}. In addition, the maximum harmonic energy ($E_{\rm max}$) in the multi-cycle regime has been shown to be linearly dependent on the electric field amplitude \cite{Luu2018, Wavelength2022}.

In remarkable contrast, the maximum energy of harmonic spectra (designated as cut-off energy) in \cite{alexander:2022} attributed to liquid-phase isopropanol, obtained in the few-cycle regime, have been reported to be linearly dependent on the laser intensity with cut-off energies as high as $\sim$50~eV. Both properties are reminiscent of gas-phase HHG on one hand, and contrasting with previous work on liquid-phase HHG \cite{Luu2018, yin2020,svoboda2021polarization,Wavelength2022} on the other. These observations have led the authors of Ref. \cite{alexander:2022} to the conclusion that electron-scattering cross sections of liquid isopropanol were significantly reduced compared to the isolated molecule. However, in an earlier work by some authors of the presented work utilizing few-cycle pulses of a 800 nm driver, an extension of the cut-off energy was not observed \cite{yin2020}.These controversies further highlight the need for additional systematic studies of HHG in the liquid phase in varying laser conditions.

In this work we demonstrate the influence of pulse duration of liquid-phase HHG by presenting back-to-back measurements of liquid- and gas-phase high harmonics in the sub-two cycle regime of 1.8 $\mu m$ laser wavelength. Our results clearly show that the cut-off energy ($E_{\rm c}$) remains pulse-width independent with a very weak dependence on intensity(if any) in the case of the liquid-phase high-harmonic spectrum throughout the transition from the multi-cycle ($\sim 50$ fs) to the sub-two-cycle regime ($\sim 11.5$ fs) regime. In comparison, the gas-phase harmonics show the expected linear dependence on the laser intensity for both $E_{\rm c}$ and $E_{\rm max}$. We also demonstrate the pulse-width independence of $E_{\rm c}$ to be a general property of harmonic spectrum generated from ethanol, isopropanol and heavy water (D$_2$O). 

Our work provides a novel, and completely independent confirmation of the fact that $E_{\rm c}$ in liquid-phase high-harmonic spectra is a fundamental property of the liquids, limited by the electron mean-free paths, and is weakly influenced by laser parameters, such as wavelength, pulse duration or intensity. These observations are confirmed by ab-initio calculations of HHG in liquids \cite{Ofer2022}. Consistent with the scattering model introduced in Ref. \cite{Wavelength2022}, the experimental and numerical results demonstrate that even in the few-cycle regime, the experimental results can be fully explained in terms of electron scattering as the limiting factor determining $E_{\rm c}$ for harmonics generated in the liquid phase. In addition, as inferred from the continuity of the harmonic spectra, the measurement suggests liquid-HHG as source of ultrashort extreme-ultraviolet pulses at lower threshold intensities in comparison to the gas phase.\\

\onecolumngrid

\begin{figure}[ht!]
\includegraphics[width=0.8\textwidth]{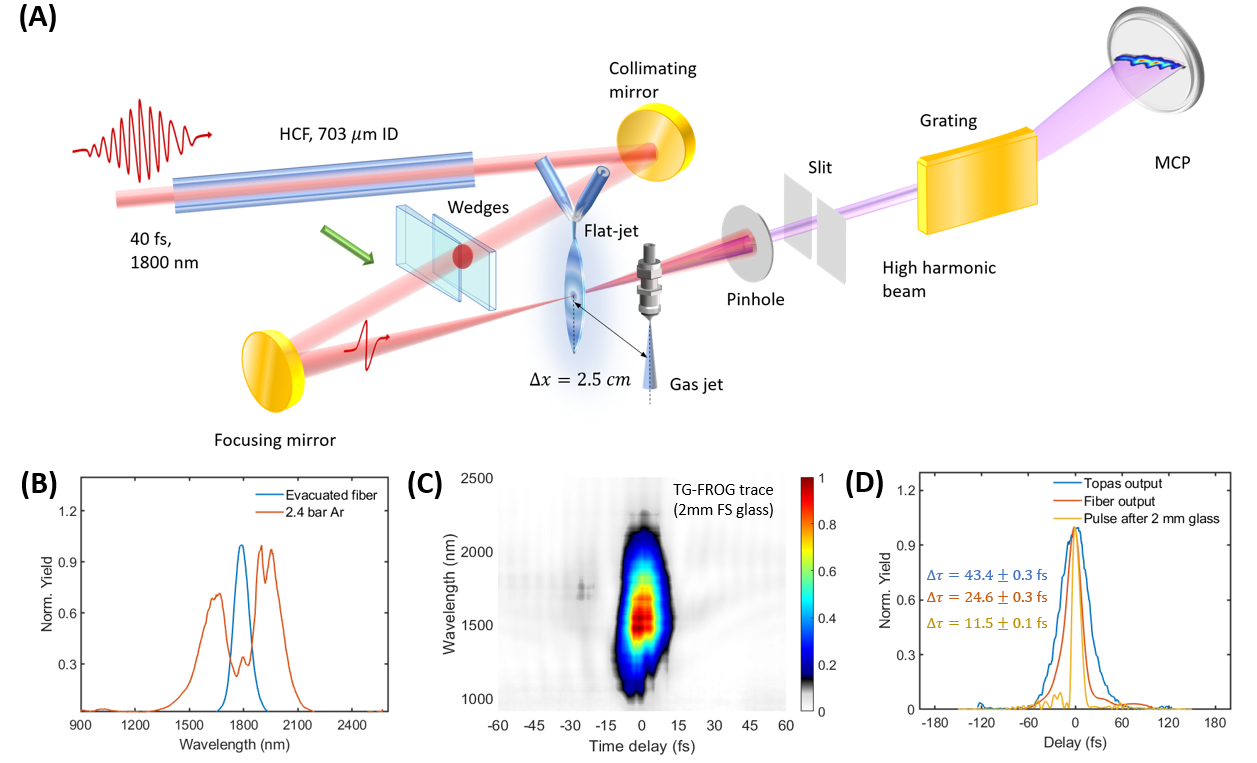}
\caption{(A) Schematic of the experimental setup. Laser pulses with a central wavelength of 1800 nm, pulse energy of 800 $\mu$J and pulse width of $\sim$43 fs are focused into a HCF of 703 $\mu$m inner diameter filled with 2.4 bar Ar. The exiting broadened pulse passes through a 2-mm fused-silica window before being focused on the flat-jet target for generating high harmonics. The harmonics pass through a slit into the XUV spectrometer that diffracts the different orders onto an MCP backed with a phosphor screen. The gas-phase sample is delivered into the laser beam through a heatable bubbler coupled to a nozzle. This setup is mounted on the same 3D manipulator as the liquid jet, such that back-to-back measurements of liquid- and gas-phase samples can be realized by a lateral translation of 2.5 cm.
(B) A comparison of the unbroadened IR spectrum obtained from an evacuated fiber (blue line) and the broadened IR spectrum in presence of 2.4 bar Ar (orange line).
(C) The transient-grating FROG trace of the compressed laser pulses. 
(D) Comparison of FWHM pulse-widths measured using FROG for the initial TOPAS output (blue line), at the output of the fiber (orange line) and after the 2~mm FS window (yellow line).FWHM widths obtained from Gaussian fits are indicated in the plot.\\}
\label{fig:one}
\end{figure}

\twocolumngrid

\section{Experimental Setup}
Figure~\ref{fig:one}(A) shows a schematic of the experimental setup. A commercial 0.8 $\mu$m, kHz Ti-Sapphire laser coupled with an optical parametric amplifier HE-TOPAS is used to generate 800 $\mu$J laser pulses centered at 1.8 $\mu$m with a duration of 43 fs (blue trace in Figure~\ref{fig:one} (D)). These pulses are then coupled into a hollow-core fiber (HCF) of 703 $\mu$m inner diameter with a 1.5 m focal length lens. The HCF is filled with 2.4 bar of Ar to generate a broadened optical spectrum as shown by the orange trace in Figure~\ref{fig:one}(B). A pulse width of $\sim 25$ fs is obtained from the output of the HCF (orange trace in Figure~\ref{fig:one} (D)). Further compression is provided by the 2 mm fused silica window of the experimental chamber. This results in a compressed pulse of $\sim$11.5 fs full-width at half maximum (FWHM) as shown by the yellow trace in Figure~\ref{fig:one}(D), using a Transient-Grating Frequency-Resolved Optical Gating (TG-FROG) \cite{Sweetser:97} measurement (Figure~\ref{fig:one}(C)). 

Figure~\ref{fig:one} (B) shows that the IR spectrum for the case of an evacuated fiber is unbroadened, corresponding to a multi-cycle pulse (for an evacuated fiber) with a measured pulse duration of $\sim 43$ fs, i.e. a Fourier-limited pulse. The harmonic spectrum generated from such a multi-cycle pulse is expected to consist of discrete harmonics as a consequence of the periodicity of the HHG process. As one approaches the single-cycle regime the discrete harmonic spectrum is expected to evolve into a continuous one \cite{Schmidt_2012}.
These pulses are then focused on a liquid flat-jet of $\sim 1 \mu$m thickness with a concave mirror of 40~cm focal length. The generated harmonics are diffracted by a XUV grating onto a multi-channel plate (MCP) with a phosphor screen and optical CCD camera for detection. 
The flat-jet, formed by the colliding jet geometry \cite{Luu2018} is mounted on an XYZ manipulator for finer adjustment with respect to the laser beam. For a back-to-back measurement of the gas-phase harmonics a heatable bubbler is mounted on the same XYZ stage at a distance of 2.5~cm from the flat-jet, which is five times the spatial lateral extension of the fat jet. We can therefore measure the gas-phase harmonics by simple translation along the lateral direction. The MCP is maintained at a voltage of -1.6 kV and the phosphor at 3.3 kV for detection of the liquid-phase harmonic spectra with an acquisition time of 200 ms. Each data set is then averaged over 30 such spectra. As the signal yield from the gas-phase harmonics is lower, the MCP voltage is increased to -1.7 kV with an acquisition time of 600 ms and 30 spectra are averaged for the gas-phase data. 

\onecolumngrid

\begin{figure}[htbp!]
\includegraphics[width=0.8\textwidth]{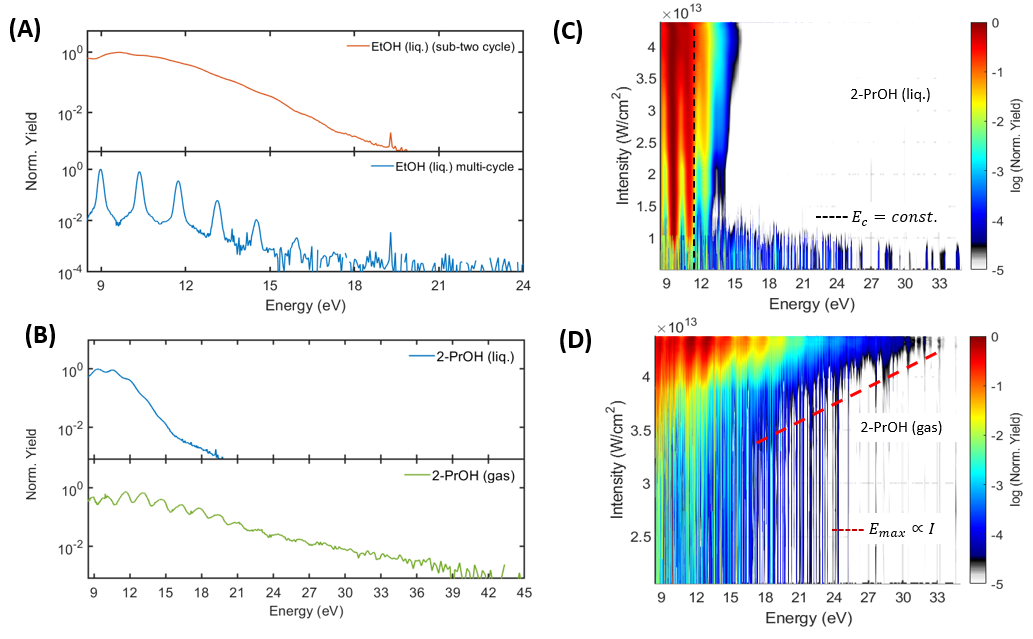}
\caption{(A) A comparison of the high-harmonic spectrum from the uncompressed laser pulse (IR beam passing through evacuated fiber) (blue line) and the compressed sub-two cycle laser pulse (IR beam passing through the fiber filled with 2.4 bar Ar, orange line). 
(B) A comparison of the harmonic spectrum generated from liquid-phase isopropanol (blue line) and gas-phase isopropanol (green line) using sub-two cycle 1.8$\mu$m, $\sim$ 12 fs laser pulses. (C) Harmonic spectra from liquid-phase isopropanol for a range of intensities driven by sub-two cycle pulses. 
(D) Harmonic spectra from gas-phase isopropanol for a range of intensities driven by sub-two cycle pulses. Each harmonic spectrum (both for liquid and gas phase) is normalized to its maximal signal intensity. The maximal intensity is limited by the onset of plasma generation in the flat-jet target.\\}
\label{fig:two}
\end{figure}

\twocolumngrid

\section{Results and discussions}

\subsection{Observation of high-harmonic generation from bulk liquids using sub-two cycle pulses}

Figure~\ref{fig:two}(A) shows a comparison of the harmonic spectrum obtained from the multi-cycle pulse (blue line) and from a sub-two cycle pulse (orange line), where a clear transformation of the discrete harmonic spectrum to a continuous one is observed. For both of these spectra the laser intensity is kept below the onset of plasma generation. To measure a systematic intensity dependence of the harmonic spectra in the liquid phase it is essential to eliminate possible gas-phase contributions from the liquid harmonic spectrum. As the cut-off energies for gases scale linearly with intensity, it may introduce a pseudo linear dependence of the harmonic energy with the laser intensity. Generally, a reference gas spectrum can be obtained by simply translating the flat-jet laterally \cite{Wavelength2022, Luu2018, yin2020}. As evaporation from the flat-jet creates a gas-phase background, this method is sufficient for acquiring gas phase spectra up to 1500 nm wavelength \cite{Luu2018}. However, we observed that at 1800 nm (both for multicycle and single cycle regime) no detectable gas-phase harmonic signal were present when the laser was focused 0.5 mm (about the lateral size of the flat jet) from the center of the jet where the gas signal was measured for shorter wavelengths \cite{Wavelength2022}. This is because as we shift to longer wavelengths the photon energy reduces and the incident intensity is not sufficient to generate detectable signal from the density of the evaporating gas. This can simply be observed from the acquisition times of the signal. For harmonic spectra obtained with an 800-nm driver, a gas-phase harmonic spectrum (acquired by a lateral 0.5 mm shift from the jet center) requires an acquisition time of 200~ms, whereas similar signal intensity for the liquid-phase harmonic spectra only take 20~ms  \cite{Wavelength2022}. For the current 1800-nm measurements a collection time of 200~ms was required to obtain the liquid spectrum with decent signal-to-noise ratio. In accordance to previous measurements, this would require an acquisition time of 2000~ms for the gas-phase spectrum, which is beyond the maximum acquisition time of the system.

However, this observation reported independently for few-cycle measurements done on liquid isopropanol \cite{alexander:2022},  cannot completely rule out gas contribution in the liquid harmonic spectra. At 1800~nm and the chosen focusing conditions, the Rayleigh range amounts to $\sim 1$ cm. Therefore, at sufficiently high laser intensities, the laser beam passing through the jet has enough intensity to generate harmonics from the gas layer adjacent to the back surface of the liquid jet. The gas density indeed monotonically decreases with the distance from the liquid-gas interface. In addition, the driving laser itself causes heating of the liquid, which increases evaporation resulting in an increased gas density for the next arriving laser pulse. This effect is absent when we laterally shift the flat-jet out of the interaction region.
Therefore, it is essential to do a back-to-back measurement of both gas and liquid phases separately to exclude gas-phase contributions to the liquid-phase harmonics.

 Isopropanol was chosen as the target for the liquid and gas-phase comparative study due to its higher vapor pressure as compared to water, which makes it easier to form a denser gas jet. Figure~\ref{fig:two}(B) shows a comparison of the harmonic spectrum generated from isopropanol in gas- (green line) and liquid-phase (blue line) with a sub-two-cycle 1800~nm laser pulse with a peak intensity of 4.4$\times$10$^{13}$W/cm$^2$. Since the high-harmonic signal from gases is dependent on the position of the target with respect to the laser focus \cite{Lewenstein2015,Steingrube2009}, the comparative study was performed at a position where the maximum signal intensity of gas-phase harmonics was observed. As expected, we observe that the liquid $E_{\rm c}$ of $\sim 11.3$ eV(determined using the technique elaborated in section 3.3 consistent with the Lewenstein definition\cite{Lewenstein1994}). Further, it is observed that the maximum photon energy for the gas-phase spectrum extends up to 35~eV, whereby the detection limit is determined by the background signal.

\begin{figure}[h]
\includegraphics[width=0.8\columnwidth]{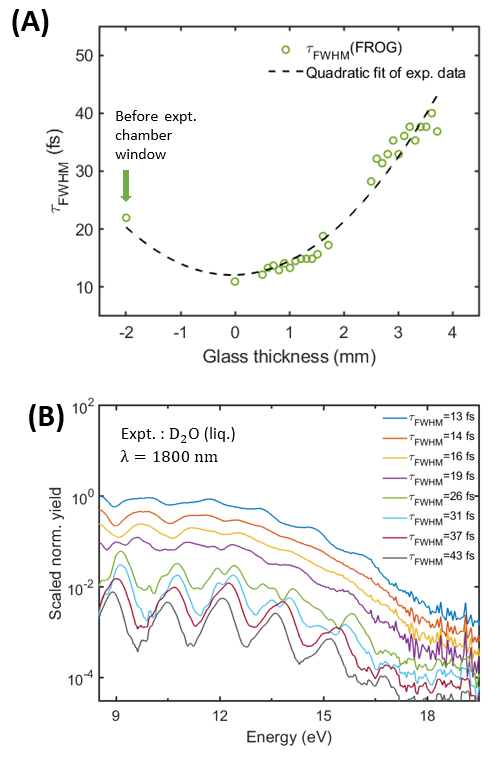}
\caption{ (A) Quadratic fit of experimentally measured FROG data to calibrate pulse width (FWHM) as a function of glass thickness. The data points at -2 mm glass thickness correspond to the FROG trace measurement at the output of the HCF before transmission through the 2~mm FS experimental chamber window. (B) Harmonic spectra from liquid-phase D$_{2}$O for different glass thicknesses. The normalized harmonic spectra at different pulsewidths have been scaled down by factors of  1/2 (14 fs), 1/4 (16 fs), 1/8 (19 fs), 1/16 (26 fs), 1/32 (31 fs), 1/64 (37 fs) and 1/128 (43 fs) respectively.The data sets were taken at a fluence of $\sim$ 0.51 J/cm$^2$.}
\label{fig:three}
\end{figure}
 The incident laser intensity is varied with the help of an automated iris. At each iris position the focal spot size is imaged and the transmitted beam power is measured. As the harmonic spectra are acquired at a distance of 11~mm from the focus position, the 1/e$^2$ radius is calculated using
\begin{equation}
w(z) = w_o\sqrt{1+(\frac{z}{z_R})^2}
\end{equation}
where $w_o$ is the 1/e$^2$ radius of the beam at focus, $z$ is the distance from the focus, $z_R$ is the Rayleigh length for specific  $w_o$ and $w(z)$ is the 1/e$^2$ radius of the beam at distance $z$ from the focus. For a gaussian beam 99$\%$ of the beam power is contained in an area of radius w'(z)=1.52$w(z)$. The respective intensity of each aperture diameter is calculated as 
\begin{equation}
I = \frac{2P}{\pi w'(z)^2 t}
\end{equation}
where $P$ is the total power transmitted through the iris, $t$ is the FWHM pulse duration of $\sim$12 fs. Figure~\ref{fig:two}(C) shows the normalized harmonic spectra from liquid isopropanol over a range of laser intensities. The actual intensity on the jet  should be accurate with calculated intensities from the above measurement up to a precision of 20$\%$.

It is observed that in the liquid phase harmonic generation occurs for intensities as low as 1$\times$10$^{13}$ W/cm$^2$ and the $E_{\rm c}$ is constant at 11.3~eV, independent of the incident intensity. Another interesting feature observed is that with increasing intensity the liquid harmonic spectra transition from discrete to continuous. 
In comparison, Figure~\ref{fig:two} (D) shows the normalized harmonic spectrum of gas-phase isopropanol for similar intensities. It is observed that gas-phase harmonics appear only above a threshold intensity of 4$\times$10$^{13}$ W/cm$^2$ and, as expected, $E_{\rm c}$ shows a linear dependence on the incident laser intensity. 

To observe the effect of pulse duration on the cut-off energy of liquids, a systematic variation of the pulse duration was performed using a pair of fused-silica wedges, where one wedge was fixed and the other was translated using an automated stage, which varied the amount of additional glass in the beam path from $\sim$0.51~mm to $\sim$1.72~mm. Beyond this, an additional 2~mm fused-silica window was added in the beam path to vary the glass thickness from 2.5~mm to 3.72~mm over the same range of wedge displacements. Figure~\ref{fig:three}(A) shows the pulse duration at each glass thickness is measured using a TG-FROG set-up. The black dashed line indicates the quadratic fitting of the pulse-duration as a function of glass thickness that is used for calibrating the pulse durations for Figure~\ref{fig:three}(B). Figure~\ref{fig:three}(B) shows the harmonic spectrum obtained from liquid D$_{2}$O for different pulse durations. For clarity the normalized harmonic spectra at 14 fs, 16 fs, 19 fs, 26 fs, 31 fs, 37 fs and 43 fs have been scaled down by factors of 1/2, 1/4, 1/8, 1/16, 1/32, 1/64 and 1/128 respectively.  A clear transition from a slightly modulated continuous spectrum to a sharply peaked odd-only harmonic spectrum at higher pulse duration is observed.
An interesting phenomenon is observed at intermediate pulse durations, e.g. in the green and blue curves of Fig.~\ref{fig:three}(B), corresponding to pulse durations of 26~fs and 31~fs, respectively. In these cases, the harmonic peaks display a substructure that is best visible around 12~eV. This substructure is attributed to the interference of the direct, forward-propagating emission from the bulk liquid with a replica that has been internally reflected twice before exiting the thin liquid sheet. This assignment is supported by the photon-energy intervals of the observed structure and our previous work on the subject \cite{Yin_2020}. Importantly, we find here that these substructures disappear for shorter pulse duration, resulting in the generation of truly continuous XUV spectra. This, in turn, indicates that isolated attosecond pulses with a good temporal contrast could be generated from liquids.

\subsection{Ab-initio calculations}
From the theoretical perspective, we employed the recently developed methodology for calculating the HHG response of liquids through the use of finite-sized clusters \cite{Ofer2022}. We follow the prescription in ref. \cite{Ofer2022} and use 54-molecule water clusters for calculating the nonlinear optical response, which is further averaged over 12 orientations. The cluster approach attempts to passivate the surface contribution to the harmonic response by an additional absorbing layer placed outside of the cluster, and by freezing the surface state dynamics. \\
\begin{figure}[!htb]
\includegraphics[width=0.85\columnwidth]{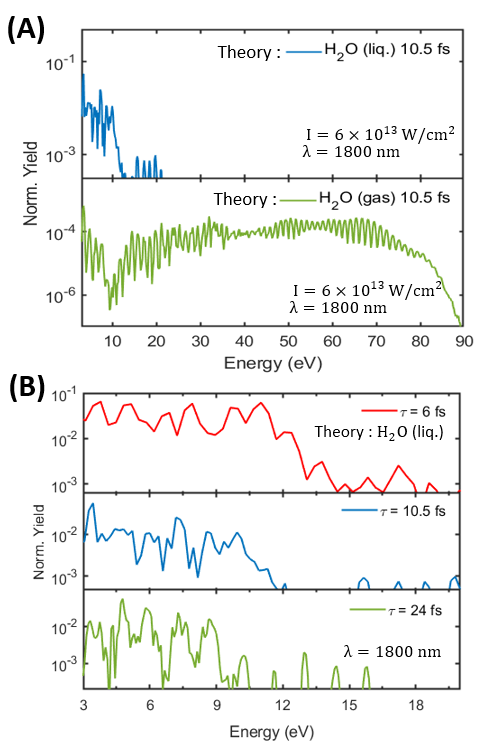}
\caption{(A) Comparison of high harmonic spectrum for liquid and gas phase H$_2$O from ab-initio calculations for 1800 nm wavelength, 10.5 fs pulse duration at an intensity of 6.3$\times$10$^{13}$ W/cm$^2$.(B) Harmonic spectra for different 6 fs, 10.5 fs and 24 fs pulse durations for liquid phase H$_2$O obtained from ab-initio calculations. The calculations were performed at a fluence of 0.66 J/cm$^2$.
}
\label{fig:four}
\end{figure}
The resulting HHG response approximately corresponds to that of the bulk liquid. We model the laser-matter interaction in the length gauge with the following electric field:\\
\begin{equation}
E(t) = E_0 f(t) \cos(\omega t +\phi_{CEP})
\end{equation}
where $E_0$ is the field amplitude, $\omega$ is taken to correspond to an 1800 nm driving wavelength, $f(t)$ is an envelope function taken as a super-sine form\cite{supsin}, which roughly corresponds to a super-Gaussian but is numerically more convenient, and $\phi_{CEP}$ is the carrier-envelope phase (CEP). We modeled several different pulse durations corresponding to different HHG regimes. For the long pulse limit we used a FWHM of 24~fs and a peak pulse intensity of 2.625$\times$10$^{13}$ W/cm$^2$, and the CEP was set to zero. The shorter pulse durations employed a peak pulse intensity that was increased linearly with respect to the pulse duration, roughly corresponding to the experimental settings. The response was further averaged over three CEPs of values of 0, $\pi/4$, and $\pi/2$, to correspond to the experimental set-up which is not CEP-stabilized. Figure~\ref{fig:four}(B) shows the numerically obtained spectra for several pulse durations, which correspond very well with the experimental results in Figure~\ref{fig:three}(B). Indeed, the cut-off in the ab-initio calculations is independent of the pulse duration, and is very weakly dependent on the pulse peak power (increasing by only ~1 eV when the peak intensity is increased by a factor of 4). Moreover, the calculations show that the harmonic peak contrast indeed decreases when employing shorter pulses, as observed experimentally. Overall, these results agree well with our previously developed semi-classical trajectory model \cite{Wavelength2022} that predicts a similar weak dependence on pulse peak power due to a suppression of longer electron trajectories through mean-free-path-limited scattering channels in the liquid. Such a strong suppression is not observed in the gas-phase calculations, which follow the expected simple three-step-model results as seen in the gas-phase simulations(bottom-panel) of Figure~\ref{fig:four} (A).

\onecolumngrid

\begin{figure}[htbp!]
\includegraphics[width=0.8\textwidth]{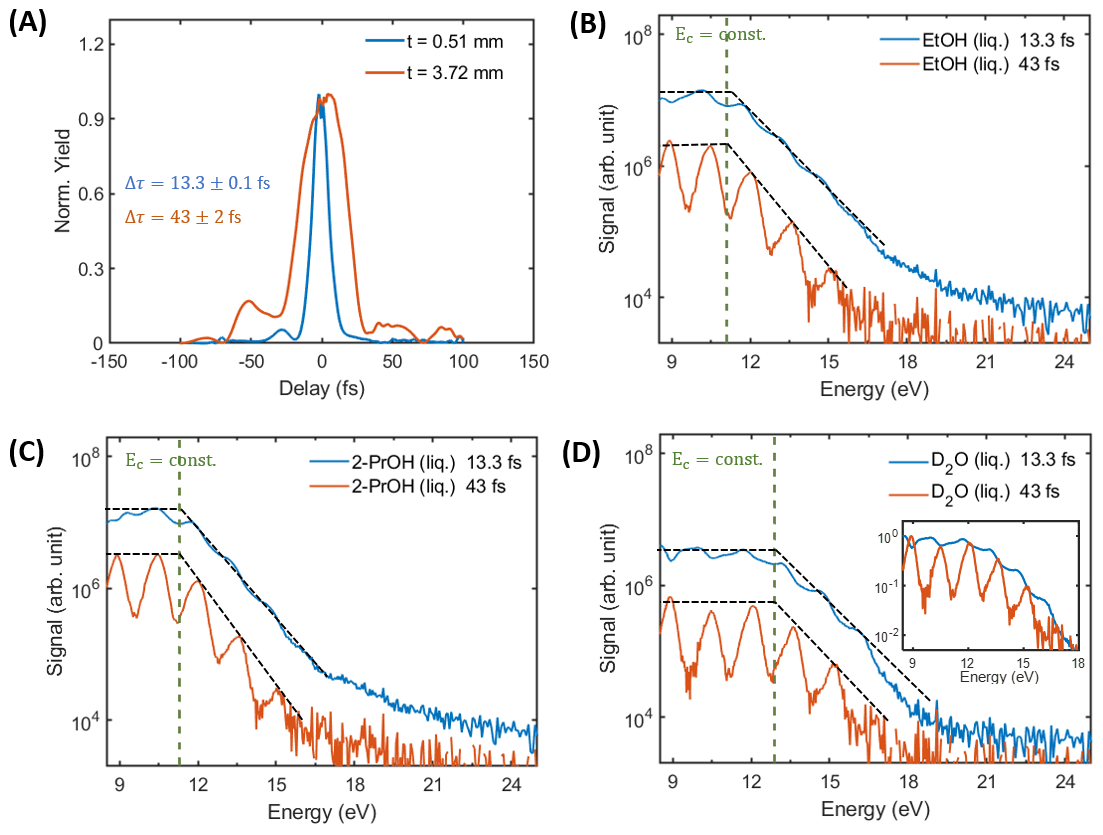}
\caption{(A) Measured temporal profiles of the sub-two-cycle pulse after it propagates through 0.51~mm FS (blue line) and 3.72~mm (orange line). FWHM widths obtained from Gaussian fits are indicated in the plot. (B-D) Harmonic spectra from liquid-phase ethanol (B), isopropanol (C) and D$_2$O (D) for different glass thicknesses Inset in Fig. 5(D) shows the same spectra as Fig. 5(D) but now normalized. The 13.3 fs spectrum is seen to envelope the 43 fs spectrum indicating negligible shift in cut-off energy. The black dased lines in all figures of Fig. 5 are guides for the eye while the green dashed lines indicate the cut-off energy determined for the multi-cycle spectrum using the fitting technique described in the manuscript.}
\label{fig:five}
\end{figure}

\twocolumngrid

\subsection{Pulse width dependence of cut-off energy for different liquids}

Finally, we also investigated whether the liquid-phase $E_{\rm c}$ changes as a function of pulse duration for different liquids. For this purpose we compare the near continuous harmonic spectrum obtained for a glass thickness of 0.51 mm with the harmonic spectrum obtained for the maximum glass thickness of 3.72 mm. Figure~\ref{fig:five}(A) shows the temporal profiles of the laser pulses measured using FROG for these two glass thicknesses at a fluence of $\sim$ 0.51 J/cm$^2$. Figure~\ref{fig:five}(B), Figure~\ref{fig:five}(C) and Figure~\ref{fig:four}(D) show that $E_{\rm c}$ is independent of pulse duration for liquid-phase ethanol, isopropanol and heavy water with their cut-off energies amounting to 11.1~eV, 11.3~eV and 12.9~eV, respectively with an accuracy of one visible harmonic order. To determine the cut-off energy (end of the plateau region) accurately in the discrete multi-cycle liquid harmonic spectrum and reduce human errors we use the a fitting approach elaborated in the supplement Figure S2. of \cite{Wavelength2022}. In brief, we explain the methodology as follows.\\

 A typical harmonic spectrum(beyond the perturbative regime) for a multi-cycle laser pulse includes plateau region with harmonic peaks of comparable signal strength followed by a cut-off region where the harmonic yield falls exponentially as a function of harmonic order. This is observed as a linear decline in a log-linear scale as shown in Figure~\ref{fig:five}(B-D). We perform a linear fitting of the log of the signal at each harmonic peak in the cut-off region as a function of the harmonic energy(represented by the black-slanted dashed lines in the orange curves of Figure~\ref{fig:five}(B-D)). The black dashed lines parallel to the energy axes in the orange curves of Figure~\ref{fig:four}(B-D) represent the average intensity value of the plateau harmonics. The intersection of these two lines(the average intensity of plateau harmonics and the linear fit of the log(signal) for the cut-off harmonics), is denoted the cut-off energy(depicted by the green dashed lines in Figure~\ref{fig:five}(B-D)). However, in the few-cycle regime where the spectrum is continuous, we do not observe discrete peaks but rather a slight modulation on a continuous spectrum. Overlaying the normalized spectrum of the few cycle case with the multi-cycle case (inset of Figure~\ref{fig:five}(D)) we see that the continuous spectrum(blue line for 13.3 fs) envelopes the discrete spectrum(orange line for 43 fs) perfectly indicating negligible change in the cut-off energy.\\
We have previously argued that, in the multi-cycle regime, HHG from liquids is well explained by a scattering-limited trajectory model. Based on this trajectory picture, $E_{\rm c}$ is determined by the characteristic electron mean-free path (MFP) in a liquid, which is shown to be comparable to the electron MFP in liquid water, methanol, ethanol and isopropanol \cite{Wavelength2022}. Determining MFPs for few-eV electrons in the liquid phase is an unsolved challenge, both computationally and experimentally. As a result, reliable low-energy electron MFPs have so far only be obtained in liquid H$_{2}$O through a combination of experimental and theoretical methods \cite{schild2020,gadeyne22a}. For the alcohols, the liquid-phase MFP has been estimated by 1/($n\sigma$), where $\sigma$ is the gas-phase elastic scattering cross-section and $n$ is the number density of scattering molecules in the liquid phase. The agreement of the MFP values derived from the scattering-limited trajectory model for the multi-cycle liquid jet experiments \cite{Wavelength2022} and the estimated MFP values proved that scattering played the decisive role for determining $E_{\rm c}$ for liquids. As a result, unlike in gases $E_{\rm c}$ is observed to be wavelength- and approximately intensity-independent in the multicycle regime.  

The cut-off energies obtained for different liquids in the current experiments using the sub-two cycle pulses are also in agreement with the multi-cycle measurements done with laser wavelengths of 800 nm, 1500 nm and 1800 nm on liquid flat-jets \cite{Wavelength2022} and is demonstrated to be intensity-independent above a certain threshold intensity. Further we demonstrate that $E_{\rm c}$ is also pulsewidth independent. The observed behavior is consistent with the scattering model of HHG in liquids. Since $E_{\rm c}$ is solely dependent on the liquid MFP and independent of the intensity, the increase in pulse duration just causes the transition of the harmonic spectrum from continuous to discrete. These results further demonstrate that even in the sub-two cycle regime, scattering plays the dominant role in the HHG mechanism in liquids, and the simple theoretical description developed in Ref.\cite{Wavelength2022} is valid.

\section{Conclusions}

In this work we have demonstrated that the cut-off energy of harmonic spectra generated in the liquid phase is pulse-duration independent. The values of the cut-off energy are moreover in agreement with our recent work in the multi-cycle regime \cite{Wavelength2022}. Through a systematic variation of the pulse duration, we have observed that harmonic spectra from liquids evolve from continuous to discrete with increasing pulse duration. Furthermore, with a back-to-back measurement of gas-phase and liquid-phase isopropanol with sub-two-cycle pulses, we showed that for liquids $E_{\rm c}$ remains intensity independent as opposed to the expected linear intensity dependence in the gas-phase spectra. We also showed that the onset of HHG in bulk liquid isopropanol occurs at much lower peak intensities (1$\times$10$^{13}$W/cm$^2$) compared to the gas phase (4$\times$10$^{13}$W/cm$^2$). \\
The present results thus show that sub-two-cycle mid-infrared driving pulses do generate fully continuous XUV spectra from bulk liquids. However, unlike in gas-phase HHG, the cut-off energy is not extended, but identical to that obtained with multi-cycle drivers. Combined with our previous results demonstrating the wavelength-independence of the cut-off energy $E_{\rm c}$ in the multi-cycle regime, the present results further confirm that scattering plays a dominant role in HHG in liquids making $E_{\rm c}$ a fundamental property of the liquid, independent of laser parameters such as wavelength, intensity and pulse duration. Owing to the continuous nature of the emitted XUV spectra which enable a more accurate determination of $E_{\rm c}$, few-cycle HHG in liquids may become an accurate method for the first all-optical determination of the mean-free paths of slow electrons in liquids, which play an important role in the understanding of radiation damage in aqueous environments. 

\begin{acknowledgments}
We thank Mario Seiler, Michael Urban and Andreas Schneider for their excellent technical support. 
We acknowledge financial support from ETH Z\"{u}rich and the Swiss National Science Foundation through grant 200021-172946. 
This work is supported by the Deutsche Forschungsgemeinschaft (DFG) through the priority program QUTIF (SOLSTICE-281310551) and the Cluster of Excellence `CUI: Advanced Imaging of Matter'- EXC 2056 - project ID 390715994, Grupos Consolidados (IT1249-19), and the Max Planck - New York City Center for Non-Equilibrium Quantum Phenomena. The Flatiron Institute is a division of the Simons Foundation.
AM acknowledges the support of the InterMUST-AoW PostDoc Fellowship. ZY acknowledges financial support from an ETH Career Seed Grant No SEED-12 19-1/1-004952-00. O.N. gratefully acknowledges the generous support of a Schmidt Science Fellowship.
\end{acknowledgments}

\section*{Competing interests}
The authors declare no conflicts of interest.

\bibliography{ref,attobib}
\balance
\end{document}